\documentclass[aps,pre,preprint,showpacs]{revtex4}
\usepackage{amsmath}
\usepackage{bm}
\usepackage[dvipdfm]{graphicx}

\begin{document}

\title{Stability analysis of twist grain boundaries in lamellar phases
of block copolymers}
\author{Xusheng Zhang$^{1}$, Zhi-Feng Huang$^{1,2}$, Jorge Vi\~nals$^{1}$}
\affiliation{$^{1}$McGill Institute for Advanced Materials, and Department of Physics,
McGill University, Montreal, QC H3A 2T8, Canada \\
$^{2}$Department of Physics and Astronomy, Wayne State University, Detroit,
MI 48201}
\date{\today }

\begin{abstract}
Twist grain boundaries are widely observed in lamellar phases of block copolymers.
A mesoscopic model of the copolymer is used to obtain stationary
configurations that include a twist grain boundary, and to analyze their
stability against long wavelength perturbations. The analysis presented is
valid in the weak segregation regime, and includes direct numerical solution
of the governing equations as well as a multiple scale analysis. We find that
a twist boundary configuration with arbitrary misorientation angle can be well
described by two modes, and obtain the equations for their slowly varying
amplitudes. The width of the boundary region is seen to scale as
$\epsilon^{-1/4}$, with $\epsilon$ being the dimensionless distance to the
order-disorder transition. We finally present the results of the linear
stability analysis of the planar boundary.

\end{abstract}

\maketitle

\section{Introduction}
\label{sec:introduction}

Block copolymers are being explored for either direct use or as processing
templates in nanolithography, photonic devices, high density storage
systems, or drug delivery, just to name a few examples
\cite{re:darling07,re:park03,re:black05,re:yoon06}. Self assembly into
mesophases of different symmetries and of controllable periodicity
(typically at the nanoscale) makes these materials a very versatile tool, and
hence the interest in studying their architectures and
self assembly mechanisms \cite{re:darling07,re:park03}. A common practical
limitation to widespread use, however, is the considerable difficulty
encountered in producing well ordered microstructures 
\cite{re:harrison00b,re:kim03,re:kramer05,re:black05}. Given that the
longest relaxation times of partially ordered microstructures are often
controlled by existing topological defects, much attention has been paid to
the motion of disclinations \cite{re:harrison00b,re:harrison04} and grain
boundaries in lamellar \cite{re:boyer02,re:huang07} and cylindrical
phases \cite{re:boyer02b}.

Theoretical analyses of defect motion have been based on a mesoscopic
description of a copolymer melt which is valid for characteristic time
scales much longer than the slowest relaxation time of the polymer chain 
\cite{re:leibler80,re:ohta86,re:fredrickson94}. Asymptotic methods commonly
employed in studies of defect dynamics in systems outside of equilibrium 
\cite{re:cross93} have been applied, to tilt grain boundaries in lamellar phases
\cite{re:tesauro87,re:manneville90}. This type of  boundary separates two
domains of differently oriented lamellae such that the plane formed by
lamellar normals of the two domains is perpendicular to the boundary plane.
Examples include boundary migration induced by lamellar curvature 
\cite{re:boyer01,re:boyer01b,re:boyer02} and the effect of an imposed shear flow
\cite{re:huang03,re:huang04}.

\begin{figure}[tbp]
\includegraphics[width=4in]{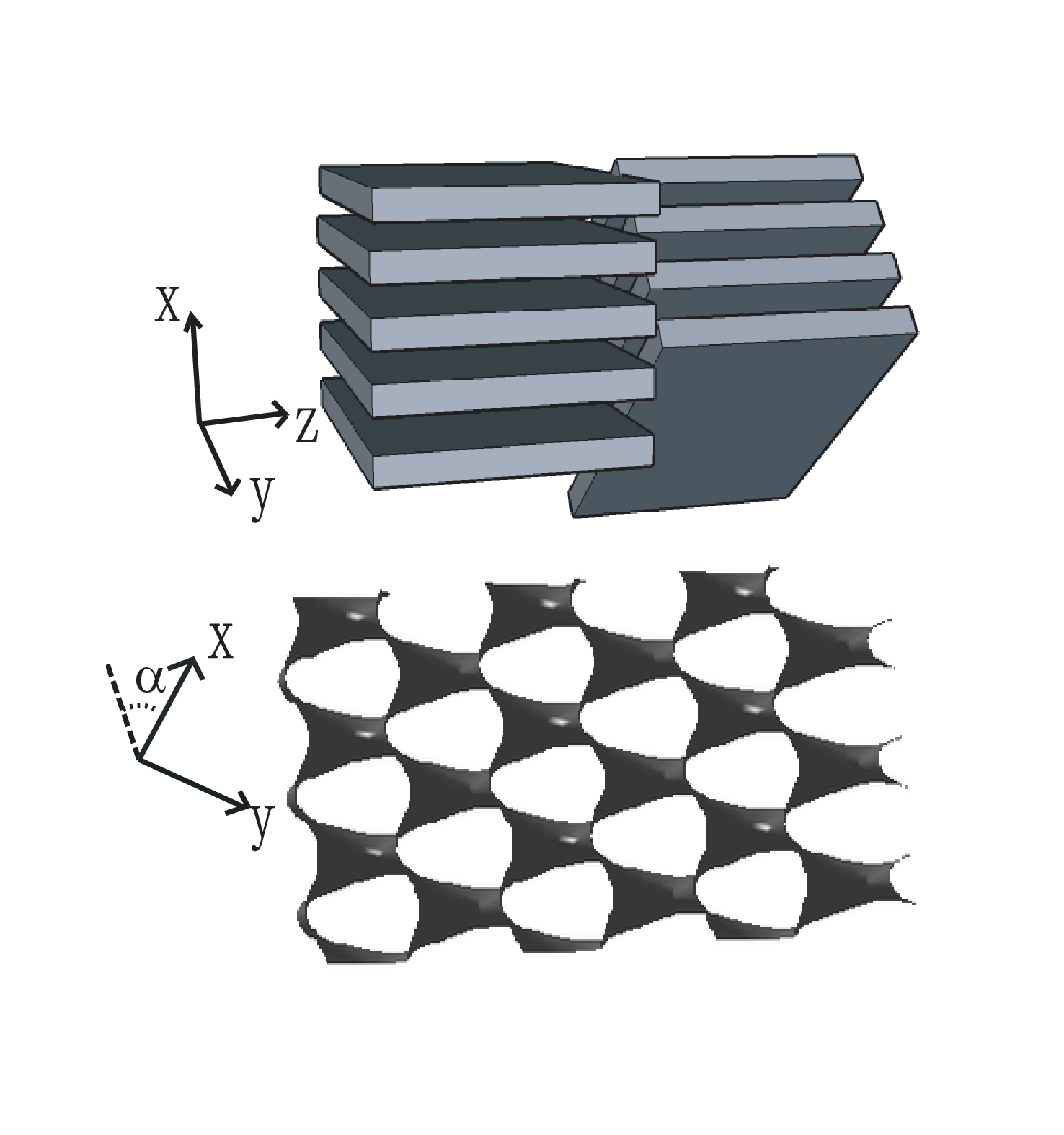}
\caption{Schematic of a twist grain boundary separating two
lamellar domains with a misorientation angle $\protect\alpha$. Also
shown (bottom) is the morphology at the boundary as given by the order
parameter of the model.}
\label{fig:tgbske}
\end{figure}

In three-dimensional samples, 90$^{\circ}$ tilt boundary
configurations (the so called T-junctions) are rarely observed in experiments
\cite{re:gido94}, possibly because they are generically unstable
\cite{re:huang05}. On the other hand, twist boundaries (such that the
wavevectors of both adjacent lamellar domains lie on the boundary plane
(see Fig. \ref{fig:tgbske})) of various
misorientation angles are commonly observed
\cite{re:thomas88,re:gido93,re:gido94_twist}. Nevertheless,
analyses of their structure and stability are still very limited 
\cite{re:kamien99,re:duque00,re:kyrylyuk05}.

We focus in this paper on a coarse grained model of a twist
grain boundary, leading to the associated amplitude (or
envelope) equation description. We obtain a stationary profile comprising a
twist grain boundary, and numerically compute its linear
stability. Our results are based on the Leibler or Swift-Hohenberg model
\cite{re:swift77,re:leibler80}, valid in the limit of weak segregation.
The analysis is conducted for a boundary of arbitrary
misorientation angle $\alpha$. In contrast with the results obtained for
the case of tilt grain boundaries, we find that the twist boundary
width scales as $\epsilon ^{-1/4}$, with $\epsilon $ the dimensionless
distance to the order-disorder threshold, and that the twist boundary is
linearly stable to long wavelength modulations for any angle $\alpha $,
consistent with experimental findings in copolymer melts.

\section{Model}
\label{sec:model}

\subsection{Coarse-grained model equation}

At a mesoscopic level, a weakly segregated diblock copolymer melt close to
the order-disorder transition temperature $T_{\mathrm{ODT}}$ is
described by a free energy,
function of monomer composition, given by Leibler 
\cite{re:leibler80,re:fredrickson87}. The corresponding relaxational dynamics
leads to the Swift-Hohenberg model equation
\cite{re:swift77,re:cross93,re:fredrickson94}. For a symmetric
diblock melt (with equal volume fraction of the two constituent monomers),
this model equation is (in dimensionless units)
\begin{equation}
\frac{\partial \psi }{\partial t}=\epsilon \psi -(\nabla
^{2}+q_{0}^{2})^{2}\psi -{\psi }^{3},  
\label{SH2}
\end{equation}
where the order parameter field $\psi$ represents the local density
difference between the two monomers of the diblock, and $q_{0}=1$ after
rescaling, although we retain the symbol $q_{0}$ in what follows for clarity
of presentation. As already stated above, the control parameter $\epsilon$
measures the distance from the order-disorder transition or bifurcation
point at which $\epsilon =0$. For $\epsilon >0$ (temperature below 
$T_{\mathrm{ODT}}$), a pattern with lamellar symmetry emerges, although usually
accompanied with large amount of defects. Therefore typical configurations display
a multidomain microstructure.

\subsection{Amplitude equations of twist grain boundaries}

Following the standard multiple scale approach in the weak segregation limit
\cite{re:manneville90,re:cross93}, we can separate the fast spatial/temporal scales
of a base lamellar pattern from its slowly varying amplitude, and derive the
associated amplitude equations for a twist grain boundary. The derivation is
based on the 
method given in Ref. \cite{re:huang04}. The order parameter field $\psi$ is
expanded as the superposition of two base modes
\begin{equation}
\psi =\frac{1}{\sqrt{3}}\left[ A\exp \left( i\vec{q}_{1}\cdot \vec{r}\right)
+B\exp \left( i\vec{q}_{2}\cdot \vec{r}\right) +\mathrm{c.c.}\right] ,
\label{op}
\end{equation}
where $\vec{q}_{1}=q_{0}\hat{x}$ and $\vec{q}_{2}=q_{0}\left( \cos \alpha
\hat{x}+\sin \alpha ~\hat{y}\right) $ (with $\alpha $ the twist angle) are
the orientations of two domains adjacent the twist
boundary (see Fig. \ref{fig:tgbske}). The
evolution of the complex amplitudes $A$ and $B$ is governed by (to leading
order in $\mathcal{O}(\epsilon ^{3/2})$)
\begin{eqnarray}
\partial _{t}A &=&\left[ \epsilon -\left( {\nabla }_{\parallel
_{1}}^{2}+2iq_{0}\partial _{\vec{n}_{1}}\right) ^{2}\right]
A-|A|^{2}A-2|B|^{2}A,  \label{ampe_vector_A} \\
\partial _{t}B &=&\left[ \epsilon -\left( {\nabla }_{\parallel
_{2}}^{2}+2iq_{0}\partial _{\vec{n}_{2}}\right) ^{2}\right]
B-|B|^{2}B-2|A|^{2}B,  \label{ampe_vector_B}
\end{eqnarray}
where $\vec{n}_{1},\vec{n}_{2}$ are the normals to the lamellar planes in
domains $A$ and $B$ respectively, ${\nabla }_{\parallel _{1}}^{2}$ is the
the Laplacian operator on the lamellar plane of domain A, and 
${\nabla }_{\parallel _{2}}^{2}$ represents the Laplacian operator on the 
lamellar
plane of domain B. For instance, if $\vec{n}_{1}=\hat{x}$ (i.e. 
$\vec{q}_{1}=q_{0}\hat{x}$), ${\nabla }_{\parallel _{1}}^{2}=\partial
_{y}^{2}+\partial _{z}^{2}$ and $\partial _{\vec{n}_{1}}=\partial _{x}$.

\section{Stationary twist grain boundary configuration}

\begin{figure}
\includegraphics[width=4in]{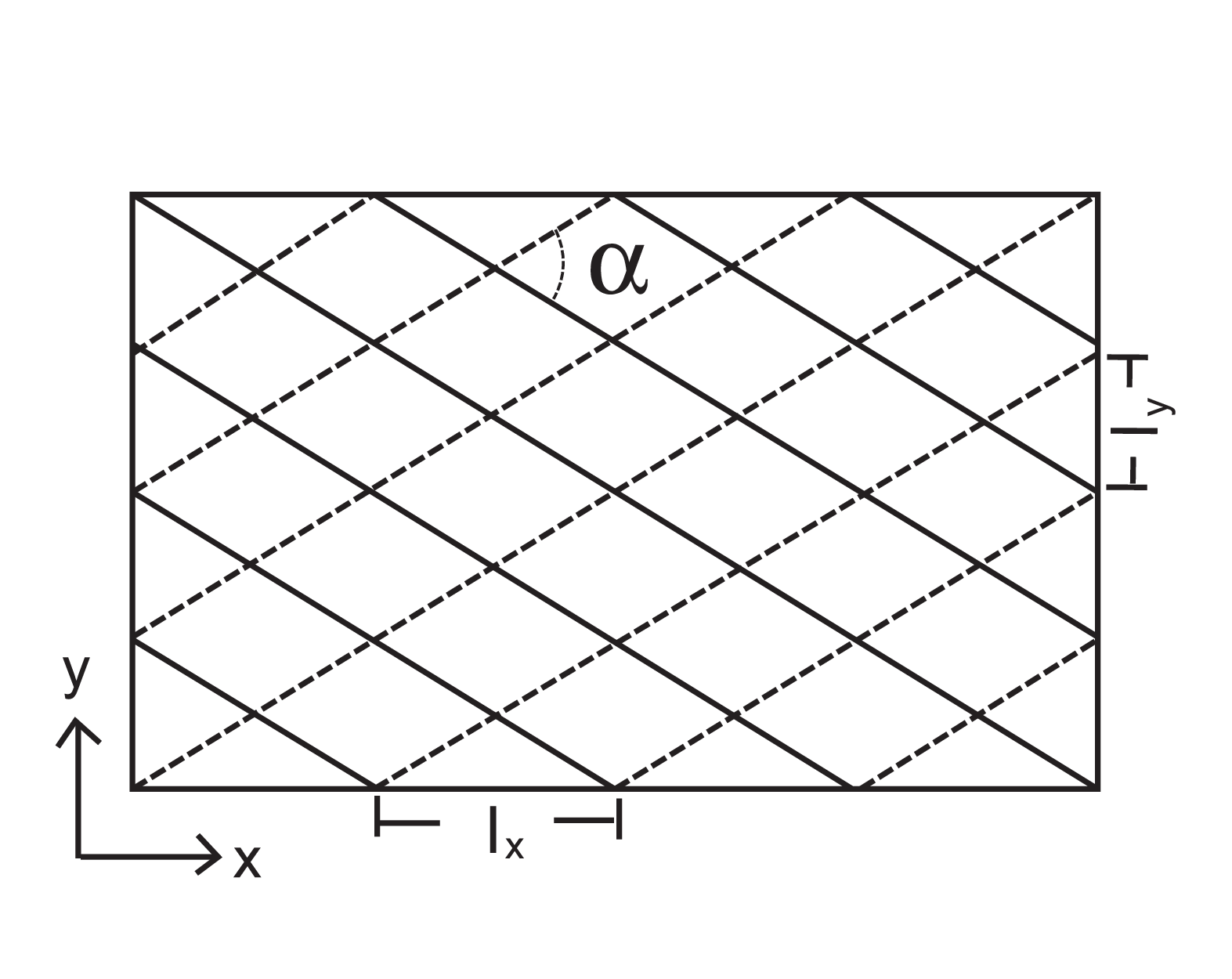}
\caption{Schematic of the constraint imposed by periodic boundary
  conditions. Solid and dashed lines represent two misoriented lamellar layers
  at the grain boundary. In order to accommodate two domains with a relative
  misorientation $\protect\alpha $ and use periodic boundary conditions, one
  needs to choose as unit cells dimensions $l_{x}=\protect\lambda _{0}/\sin
  (\protect\alpha /2)$, $l_{y}=\protect\lambda _{0}/\cos (\protect\alpha /2)$
  along the $x$ and $y$ directions on the boundary plane.}
\label{fig:grids}
\end{figure}

The steady state configuration of twist grain boundaries has been first examined
through direct numerical solution of the model equation (\ref{SH2}). A
pseudospectral method in Fourier space is adopted, with periodic boundary
conditions in all three directions. A Crank-Nicholson time stepping scheme
is applied to the linear terms, with a second order Adams-Bashford algorithm
used for the nonlinear term. Periodic boundary conditions are satisfied
through the consideration of an initial configuration comprising a symmetric
pair of twist boundaries that are sufficiently far apart so that
their motion is approximately independent. An additional restriction needs to
be placed on the dimension of the computational cell along the $x$ and $y$
directions on the plane of the grain boundary (as shown schematically in 
Fig. \ref{fig:grids}). Due to the requirement that an integer multiple of
lamellar periods $\lambda_0$ ($=2\pi / q_0$) must equal the length of the
computational cell in the direction parallel to the lamellar normal, the
unit cell lengths are $l_{x}=\lambda _{0}/\sin (\alpha /2)$
and $l_{y}=\lambda _{0}/\cos (\alpha /2)$. We consider a uniform spatial
discretization in a $L_x \times L_y \times L_z$ grid with spacings $\Delta
x=l_{x}/16$, $\Delta y=l_{y}/16$, and $\Delta z=\lambda _{0}/16$,
corresponding to 16 grid points per unit cell length. Most calculations
shown below correspond to a system size of $256^3$, with a dimensionless
time step $\Delta {t}=0.2$ used in the numerical integration.

\begin{figure}
\includegraphics[height=2.5in]{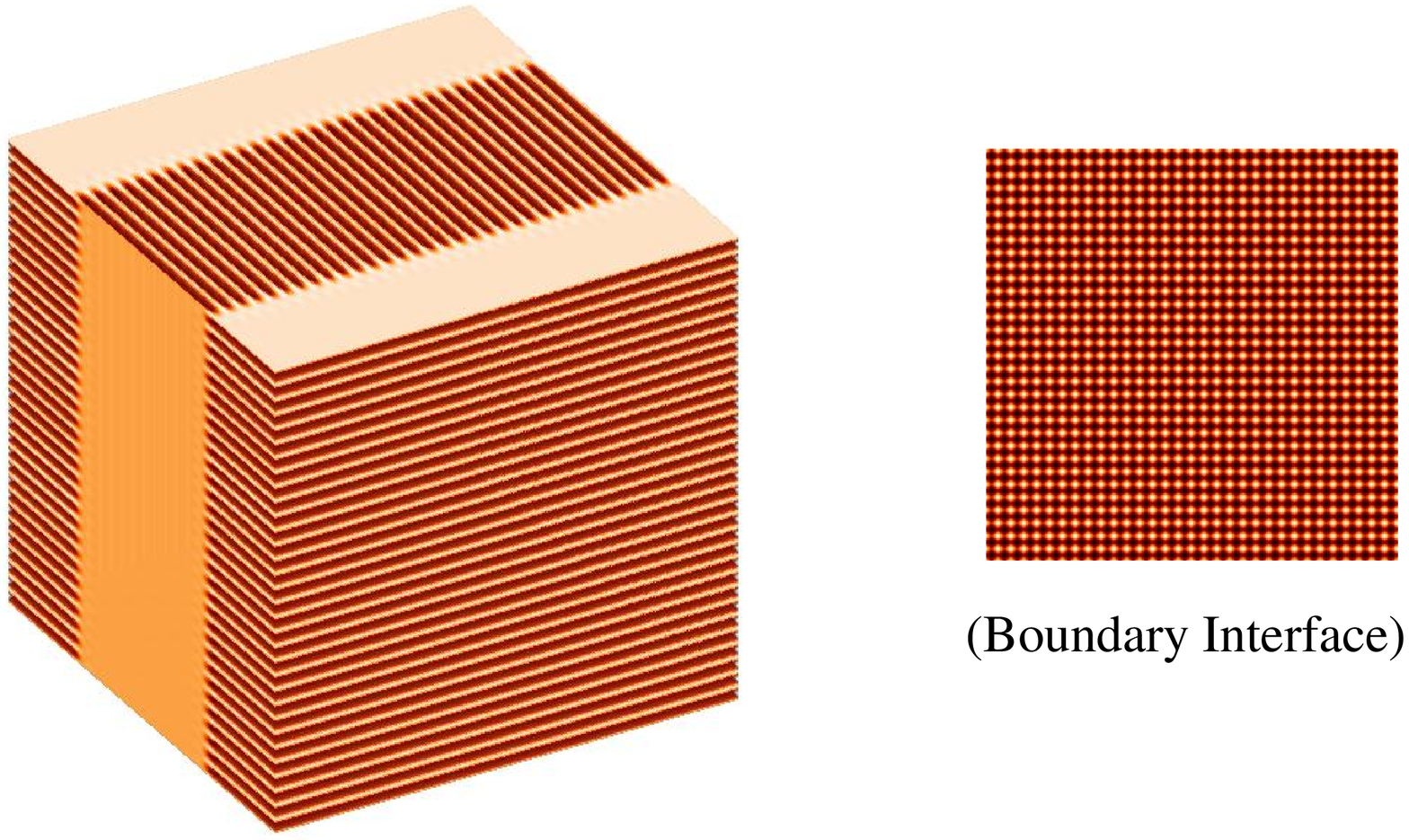}
\caption{A stationary $\alpha = 90^{\circ}$ twist grain boundary configuration in
a lamellar phase, as given by numerical solution of the model equation
(\protect\ref{SH2}). The grid size is $256^3$, and $\protect\epsilon=0.04$. Right
panel: order parameter (gray scale) at the boundary interface. }
\label{fig:twist}
\end{figure}

A typical stationary configuration is shown in Fig. \ref{fig:twist},
corresponding to $\alpha = 90^{\circ }$ with $\epsilon =0.04$ and at a
time $t=10^{4}$. Also shown in the figure (in grey scale) is the two
dimensional order parameter  at the boundary interface. It is doubly periodic
along the two directions defined by the bulk lamellar domains adjacent to the grain
boundary.

\begin{figure}[tbp]
\includegraphics[width=4in]{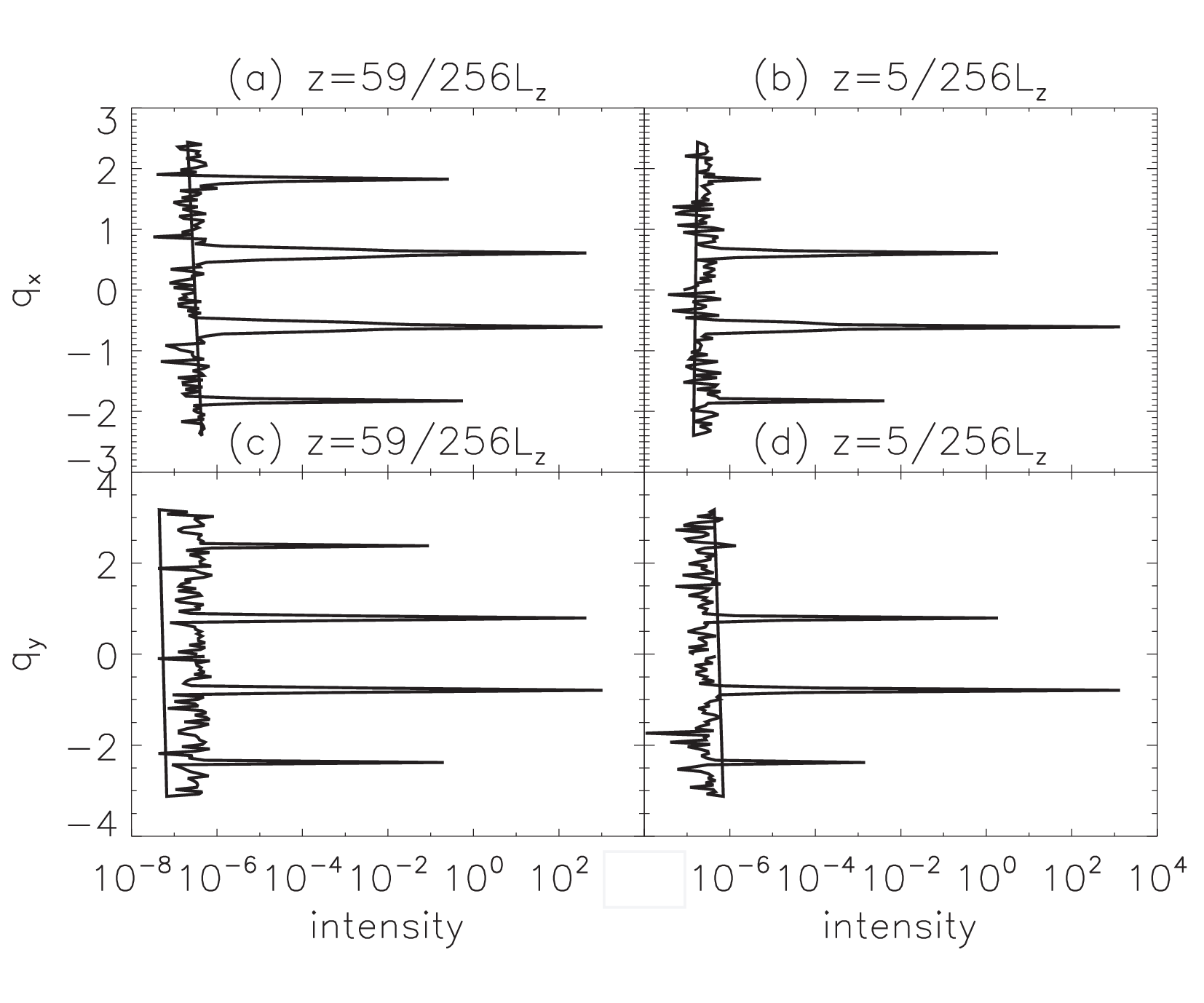}
\caption{Power spectrum of $\psi$ along specific
directions in $\vec{q}=(q_{x},q_{y})$ space, and at constant location $z$ for
a $\alpha = 75^{\circ }$ twist boundary. We show the spectrum for
$\protect\epsilon = 0.02$ and $t=2000$. Panel (a) shows the power spectrum as
a function of $q_{x}$ for $q_{y}=q_{y0}$ and $z=(59/256)L_{z}$, whereas (b)
shows the spectrum far into the bulk at $z=(5/256)L_{z}$.
Panels (c) and (d) show similar spectra as a function of $q_{y}$
for $q_{x}=q_{x0}$ and at $z=(59/256)L_{z}$ (c) or $(5/256)L_{z}$ (d). Here 
$(q_{x0},q_{y0})$ is the location of the peak of the two dimensional
power spectrum for the order parameter $\protect\psi$.}
\label{fig:fspectrum}
\end{figure}

Before we carry out a multiple scale analysis, we have checked the underlying
assumption that the stationary order parameter field $\psi$ can be decomposed
into two Fourier modes (see Eq. (\ref{op})). We calculate the Fourier spectrum
of the order parameter field both at the grain boundary and in the bulk. We
illustrate our findings with the case of a $\alpha = 75^{\circ }$ twist
boundary with $\epsilon =0.02$. The two dimensional Fourier spectrum of $\psi$
on the boundary plane ($z=(59/256)L_{z}$) shows four maxima at wavevectors $(\pm
q_{x0},\pm q_{y0})$, with $\sqrt{q_{x0}^{2}+q_{y0}^{2}}=q_{0}$ ($q_{0}$ is the
wavenumber in the bulk). We also observe $2\arctan (q_{x0}/q_{y0})=75^{\circ}$
(exactly the misorientation angle $\alpha$). This reflects the fact that the
order parameter in the grain boundary region is a combination of the two bulk modes.
The same conclusion is supported by an analysis of
higher harmonics in the spectrum. Figures \ref{fig:fspectrum}a and 
\ref{fig:fspectrum}b show the intensity of the spectrum along $q_{x}$ at 
$q_{y}=q_{y0}$ and at two different values of $z$: one within the grain
boundary (Fig. \ref{fig:fspectrum}a), the other in the bulk A phase
(Fig. \ref{fig:fspectrum}b) (identical conclusions can be drawn from the
analysis in phase B). Figures \ref{fig:fspectrum}c and \ref{fig:fspectrum}d
show the same quantity but as a function of $q_{y}$ at $q_{x}=q_{x0}$, and
for the same two values of $z$. The fact that all the visible harmonics
within the grain boundary region are almost the same as in the bulk suggests
that in the weak segregation limit considered here, the superposition
of the two bulk modes in Eq.~(\ref{op}) used for the multiple scale analysis
that will follow appears to be sufficient for the description of the order
parameter profile around the grain boundary.

\section{Asymptotic boundary width}

The boundary width $\delta$ of a twist grain boundary as a function of
$\epsilon$ in the limit $\epsilon \rightarrow 0$  can be determined either
numerically from the stationary configuration given above, or via a multiple 
scale analysis (\ref{ampe_vector_A}) and (\ref{ampe_vector_B}). In the latter
case, simple dimensional analysis of Eqs. (\ref{ampe_vector_A}) and
(\ref{ampe_vector_B}) along the grain boundary normal (the $z$ direction)
leads to the following result
\begin{equation}
\delta \sim \epsilon^{-1/4}.  
\label{delta_scale}
\end{equation}
This is in contrast with the known behavior for a tilt gain boundary in which
$\delta \sim \epsilon^{-1/2}$ \cite{re:tesauro87}. The latter scaling behavior
follows from the fact that there are two distinct characteristic length scales
for lamellar relaxation: One along the direction parallel to
the lamellar normal with scale $l_{\perp }\propto \epsilon ^{-1/2}$,
the other along the plane of the lamella with scale $l_{\parallel }\propto
\epsilon^{-1/4}$. For a twist grain boundary, on the other hand, the direction
($z$) normal to the boundary is parallel to the lamellar planes of both
phases, and hence it is reasonable to expect that the boundary width along $z$
scales as $\epsilon ^{-1/4}$. In summary, a twist boundary is much narrower
than a tilt boundary in the limit $\epsilon \rightarrow 0$.

\begin{figure}[tbp]
\includegraphics[width=4in]{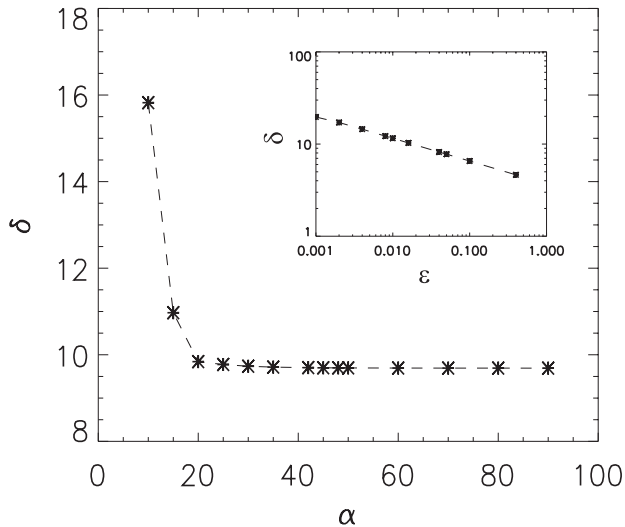}
\caption{Boundary width $\protect\delta $ (in dimensionless units)
as a function of twist angle $\protect\alpha $ for $\protect\epsilon
=0.02 $. Inset: Width $\protect\delta $ vs. $\protect\epsilon $ as obtained
from the stationary solutions of Eq. (\ref{SH2}) for $\alpha = 90^{\circ
}$. The slope of the log-log plot is $-0.244\pm 0.002$.}
\label{fig:bdwidth}
\end{figure}

The result above holds for any misorientation angle as we have verified by
numerical solution of the model equation (\ref{SH2}). We first determine the
location of the boundary by estimating the amplitude $B(z)$
\cite{re:huang04} 
\begin{equation}
B(z)\simeq \frac{\sqrt{3}}{4N}\sum_{m=1}^{N}\left[ \psi (\vec{r}\cdot \hat{n}
_{B}=m\lambda _{0};z)-\psi (\vec{r}\cdot \hat{n}_{B}=(m-1/2)\lambda _{0};z)
\right] ,
\end{equation}
with $\hat{n}_{B}$ the unit normal to lamellae $B$ and $N$ the number
of pairs of lamellae. The boundary region is chosen such that the value of
$B(z)$ lies within $10\%$ -- $90\%$ of its maximum.
Since the width of the boundary is only several times the lamellar width, a
linear interpolation algorithm is used to increase the accuracy of the
boundary location. The relation obtained between boundary width $\delta$
(in dimensionless units) and misorientation angle $\alpha$
is plotted in Fig. \ref{fig:bdwidth} for $\epsilon =0.02$.
For $\alpha >20^{\circ }$, the boundary width becomes approximately
independent of $\alpha $. Otherwise, $\delta $ increases rapidly with
decreasing angle. Although the accuracy of our numerical solution degrades
when $\alpha $ is small, the trend obtained points to a divergence of the
boundary width as $\alpha \rightarrow 0$. We find similar results when
numerically solving the corresponding amplitude equations 
(\ref{ampe_vector_A}) and (\ref{ampe_vector_B}) for small twist angles.
Figure \ref{fig:bdwidth} also shows our results for $\delta $ as a function of
$\epsilon $ for $\alpha = 90^{\circ }$. Given the spatial discretization used
in our integration, the boundary widths that we have been able to investigate
range from $3\lambda _{0}$ at $\epsilon =0.001$ to $3\lambda _{0}/4$ at $\epsilon
=0.4$ (with $\lambda_0 = 2\pi / q_0$). Within this limited range, a power law
dependence between $\delta $ and $\epsilon $ is found, with an exponent
$-0.244\pm 0.002$, in agreement with our expectation from dimensional analysis.
Analogous results have been obtained for other values of $\alpha$.

\section{Stability analysis}

As noted above, twist grain boundaries are observed in great abundance in
experiments that address the microstructure of large samples in the lamellar phase 
\cite{re:thomas88,re:gido93,re:gido94_twist}. We conduct here a linear
stability analysis of a planar boundary from the amplitude
equations (\ref{ampe_vector_A}) and (\ref{ampe_vector_B}) that are derived
from our model equation. We start from a base state involving a stationary and planar
twist boundary of arbitrary misorientation angle $\alpha$ and wavenumber
$q_{0}$ \cite{re:xz1_fo1}. The corresponding amplitudes $A^{(0)}$ and $B^{(0)}$ are assumed to
be only a function of $z$, the direction normal to the boundary, and are given by
\begin{eqnarray}
{\epsilon }A^{\left( 0\right) }-{\partial }_{z}^{4}A^{\left( 0\right)
}-|A^{\left( 0\right) }|^{2}A^{\left( 0\right) }-2|B^{\left( 0\right)
}|^{2}A^{\left( 0\right) } &=&0,  \label{A0} \\
{\epsilon }B^{\left( 0\right) }-{\partial }_{z}^{4}B^{\left( 0\right)
}-|B^{\left( 0\right) }|^{2}B^{\left( 0\right) }-2|A^{\left( 0\right)
}|^{2}B^{\left( 0\right) } &=&0.  
\label{B0}
\end{eqnarray}
We next expand the complex amplitudes around the stationary solutions
\begin{eqnarray}
A(x,y,z,t) &=&A^{\left( 0\right) }(z)+\sum_{q_{x},q_{y}}{\hat{A}
(q_{x},q_{y},z,t)e^{i(q_{x}x+q_{y}y)}}, \\
B(x,y,z,t) &=&B^{\left( 0\right) }(z)+\sum_{q_{x},q_{y}}{\hat{B}
(q_{x},q_{y},z,t)e^{i(q_{x}x+q_{y}y)}},
\end{eqnarray}
substitute these expansions into Eqs. (\ref{ampe_vector_A}) and 
(\ref{ampe_vector_B}), and linearize the resulting equations with respect to the
perturbations $\hat{A}$ and $\hat{B}$. We find
\begin{eqnarray}
\partial _{t}\hat{A}(q_{x},q_{y},z,t) &=&\left[ \epsilon -(\partial
_{z}^{2}-q_{y}^{2}-2q_{0}q_{x})^{2}-2|A^{\left( 0\right) }|^{2}-2|B^{\left(
0\right) }|^{2}\right] \hat{A}(q_{x},q_{y},z,t)  \notag \\
&&-\left( A^{\left( 0\right) }\right) ^{2}\hat{A}^{\ast
}(-q_{x},-q_{y},z,t)-2A^{\left( 0\right) }B^{\left( 0\right) \ast }\hat{B}
(q_{x},q_{y},z,t)  \notag \\
&&-2A^{\left( 0\right) }B^{\left( 0\right) }\hat{B}^{\ast
}(-q_{x},-q_{y},z,t),  \label{stabA} \\
\partial _{t}\hat{B}(q_{x},q_{y},z,t) &=&\left[ \epsilon -(\partial
_{z}^{2}-q_{y_{2}}^{2}-2q_{0}q_{x_{2}})^{2}-2|A^{\left( 0\right)
}|^{2}-2|B^{\left( 0\right) }|^{2}\right] \hat{B}(q_{x},q_{y},z,t)  \notag \\
&&-\left( B^{\left( 0\right) }\right) ^{2}\hat{B}^{\ast
}(-q_{x},-q_{y},z,t)-2B^{\left( 0\right) }A^{\left( 0\right) \ast }\hat{A}
(q_{x},q_{y},z,t)  \notag \\
&&-2A^{\left( 0\right) }B^{\left( 0\right) }\hat{A}^{\ast
}(-q_{x},-q_{y},z,t),  \label{stabB}
\end{eqnarray}
where $q_{x_{2}}=\cos \alpha ~q_{x}+\sin \alpha ~q_{y}$ and $q_{y_{2}}=-\sin
\alpha ~q_{x}+\cos \alpha ~q_{y}$.

Since we do not have an analytic expression for the amplitudes of the base
state, we study its stability by examining the temporal evolution of small
random perturbations to both real and imaginary parts of
$\hat{A}$ and $\hat{B}$ for a range of values of $q_{x}$ and $q_{y}$,
and integrating the system of Eqs.~(\ref{A0})--(\ref{stabB})
numerically. The details of the numerical algorithm and procedure are
given in Ref. \cite{re:huang05}. The parameters chosen here are $\Delta
z=\lambda_{0}/8$ for the discretization along the $z$ direction, with
$L_{z}=1024$ grid nodes (or equivalently a length of the computational domain
of $128\lambda _{0}$). The time step chosen is $\Delta t=0.2$. If
the planar grain boundary is stable, perturbations in $\hat{A}$ and 
$\hat{B}$ will decay in time for all wavevectors $(q_x,q_y)$; otherwise an
instability would manifest itself by an increase of these perturbations within
a certain range of wavevectors.

\begin{figure}[tbp]
\includegraphics[width=4in]{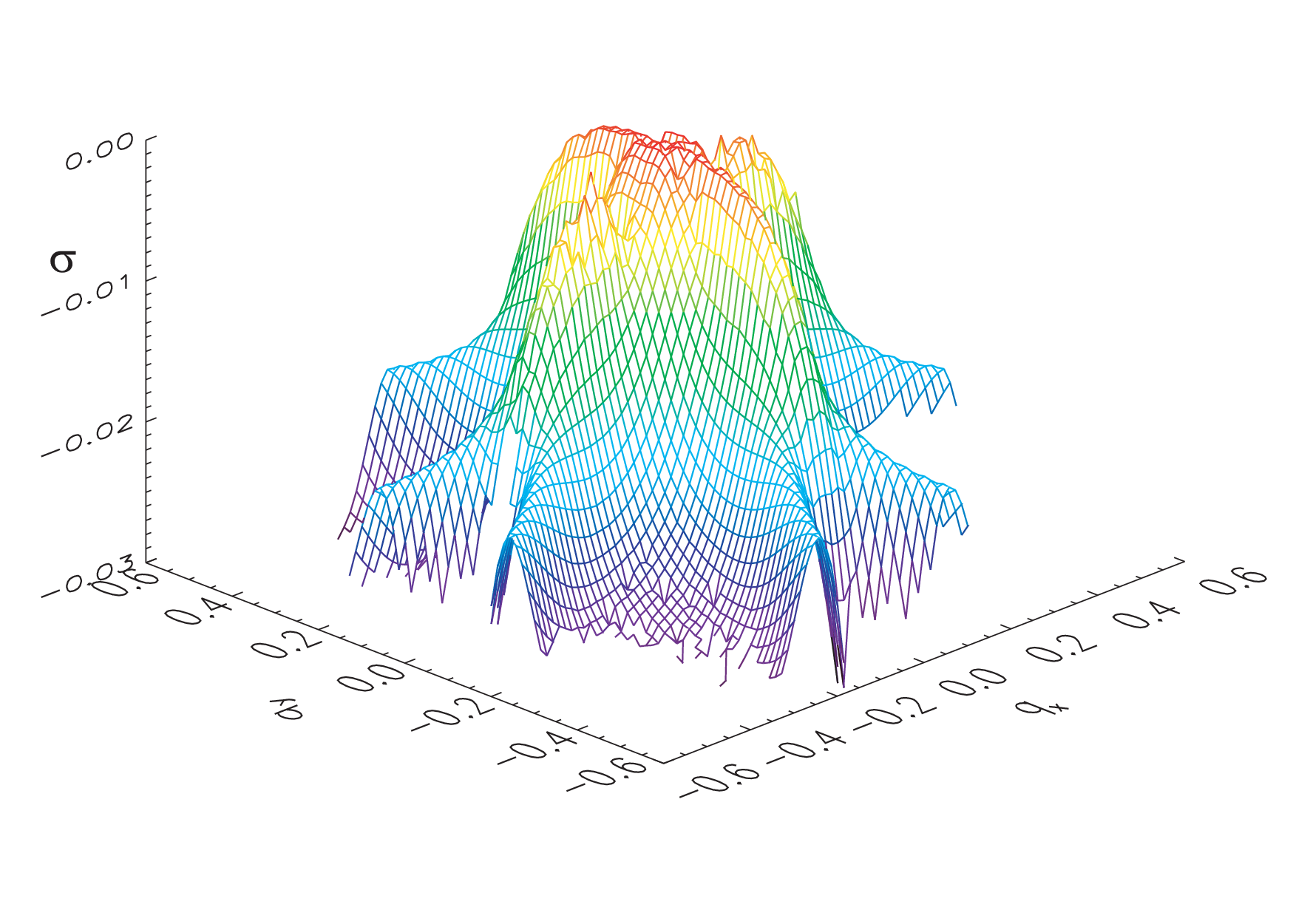}
\caption{Perturbation growth rate $\sigma $ as a function of 
wavevector ($q_{x}$, $q_{y}$) for $\epsilon =0.04$ and $\protect
\alpha =90^{\circ}$. We find that $\protect\sigma <0$ over the whole range of
wavevectors investigated.}
\label{fig:sigma_q}
\end{figure}

\begin{figure}[tbp]
\includegraphics[width=4in]{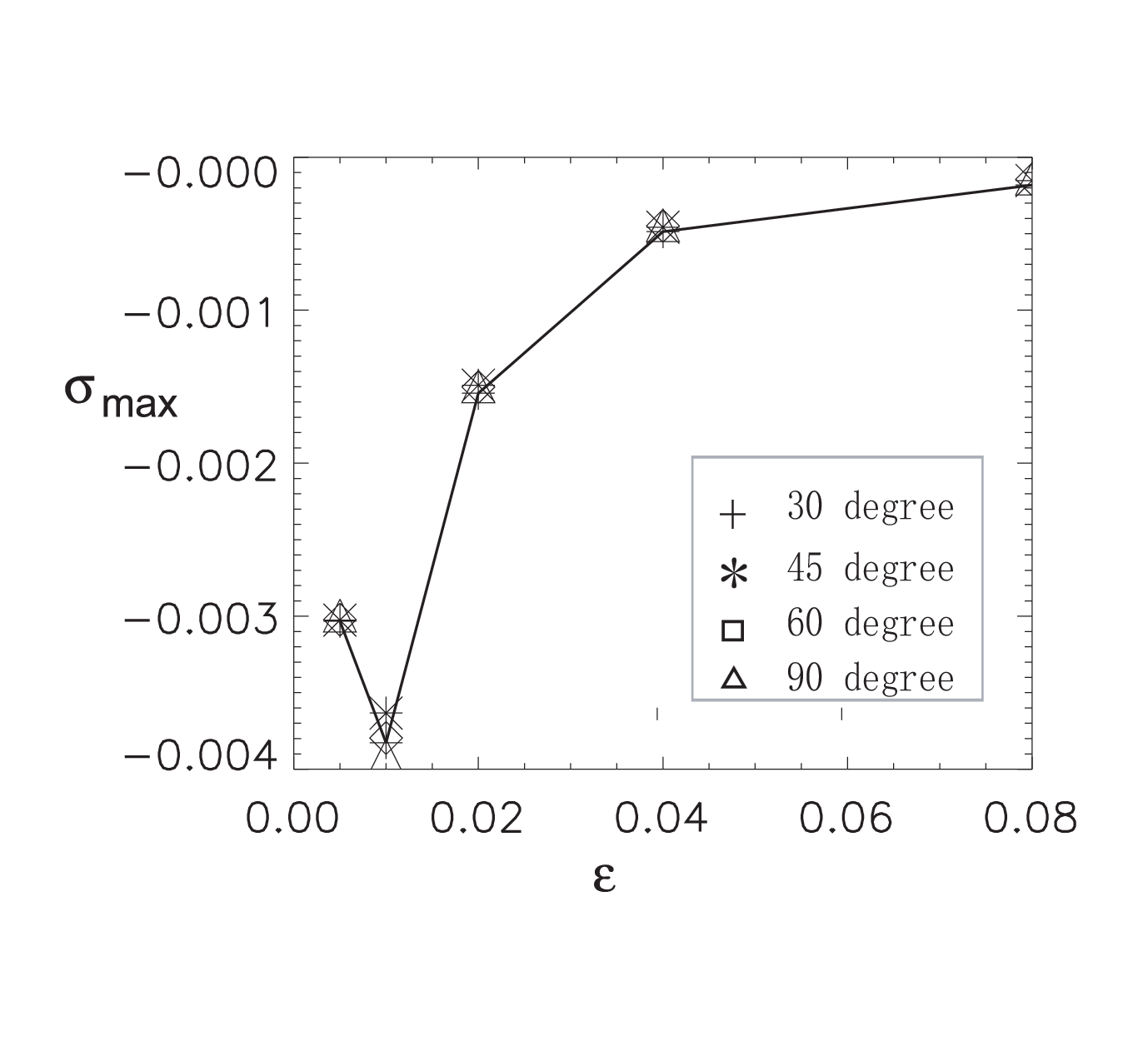}
\caption{Maximum perturbation growth rate $\protect\sigma _{\mathrm{max}}$
as a function of $\protect\epsilon $, for different twist angles 
$\alpha = 30^{\circ }$, 45$^{\circ }$, 60$^{\circ }$, and 90$^{\circ }$.}
\label{fig:epsilon_sigma}
\end{figure}

From the relaxation of the perturbations, we estimate the perturbation growth
rate $\sigma (q_{x},q_{y})$ from $|\hat{A}(t)|, |\hat{B}(t)| \propto e^{\sigma
  (q_{x},q_{y})t}$  and the numerical solutions for $\hat{A}$ and $\hat{B}$
for a given set of $(q_{x},q_{y})$. A typical result is shown in
Fig. \ref{fig:sigma_q} for $\alpha = 90^{\circ}$ and 
$\epsilon = 0.04$. We always observe that $\sigma < 0$, for all the
wavevectors of the perturbation explored. This is also the case for
different values of $\epsilon$ and angle $\alpha$, as shown in Fig.
\ref{fig:epsilon_sigma}. The maxima of $\sigma$ for $\alpha$ ranging from 
$30^{\circ}$ to $90^{\circ }$, and $\epsilon$ from 0.005 to 0.08 have been
calculated, all yielding a stable planar boundary.

\section{Conclusions}

We have used the Swift Hohenberg model as an approximate mesoscale description
for the evolution of a twist grain boundary in a lamellar phase of a diblock
copolymer.  We have shown that the order parameter field can be well
approximated in the weak segregation regime by a combination of the two modes
of the ordered lamellar phases on either side of the grain boundary. The
equations governing the slow evolution of the amplitudes or envelopes of these
modes have been derived for arbitrary misorientation angle. The stationary 
solution is only a function of the coordinate normal to the grain boundary
plane, and is characterized by a width $\delta \sim \epsilon^{-1/4}$, with
$\epsilon$ the distance from the order-disorder point. We have then conducted
a linear stability analysis by direct numerical solution of the governing
equations, and have found that the twist boundary is linearly stable within a
wide range of parameters investigated.

\begin{acknowledgments}
This research has been supported by the National Science Foundation under
grant DMR-0100903, and by NSERC Canada.
\end{acknowledgments}

\bibliography{references}

\begin{thebibliography}{32}
\bibitem[{re:()}]{re:xz1_fo1}
\bibinfo{note}{The stability analysis presented in this paper concerns only
  base states of wavenumber $q_{0}$. Twist grain boundaries separating uniform
  domains of wavenumber $q \neq q_{0}$ can be unstable. The range of stability
  as a function of $q$, at fixed $\epsilon$, is smaller than that of uniform
  domains. Therefore there is a range of wavenumbers within which the
  instability originates in the grain boundary region.}
\end{thebibliography}

\end{document}